# Implementation of a Multi-Beam MAC Protocol for Multi-Hop Wireless Networks in Riverbed Modeler

Shivam Garg[1] (Student Member, IEEE), Nandini Venkatraman[1], Soroush Tamizi[1], Hira Shah[1], Elizabeth Bentley[2] (Member, IEEE), and Sunil Kumar[1*] (Senior Member, IEEE)

[1]Department of Electrical and Computer Engineering, San Diego State University, San Diego, CA, USA
[2]Air Force Research Laboratory, Rome, NY, USA
[*] Corresponding author

{sgarg@sdsu.edu, nandinivenkatraman@yahoo.com, soroush.tamizi@gmail.com, hirajawedshah@gmail.com, elizabeth.bentley.3@us.af.mil, skumar@sdsu.edu}

***ABSTRACT:*** Recent advances in antenna technology have made the design of multi-beam antennas (MBA) feasible. Compared to an omni-directional or a single beam directional antenna, an MBA equipped node can achieve a throughput of up to *m* times, by simultaneously communicating on its *m* non-interfering beams. As a result, a few multi-beam directional medium access control (MAC) schemes have been proposed in the literature recently, which are implemented mostly on the in-house simulation setups in Matlab or C/C++. These implementations make many assumptions to simplify their design, without a thorough implementation of other network layers. However, the implementation of a multi-beam MAC scheme on the well-known discrete event network simulator platforms (such as the Riverbed Modeler, NS3, QualNet) is challenging as it requires extensive changes and additions to various source code modules. In fact, the network protocols in these simulator packages have been mainly designed for omni-directional communication, and very few implementations of directional MAC and other network protocols exist in the literature.

This paper presents a framework to implement a multi-beam directional MAC scheme in multi-hop wireless networks, by using the Wireless Suite of Riverbed Modeler. The detailed implementation procedures are described for multi-beam antenna module, multi-beam node model, concurrent packet transmission and reception, scheduling, collision avoidance, retransmission, and local node synchronization. These MAC modules and methodology can be very helpful to the researchers and developers for implementing the single-beam as well as multi-beam directional MAC and routing protocols in Riverbed Modeler.

## 1. INTRODUCTION

Existing wireless networks primarily use omni-directional communication, which gives rise to several challenges, including the interference-limited capacity, power efficiency, and quality of service. Recently, there has been increased interest in directional communication, which provides significant improvements by enabling the spatial reuse, extending coverage, mitigating interference, and thereby increasing the network capacity [1-3]. At the same time, the recent advances in antenna technology, along with the shift toward higher frequencies, have made the design and use of multi-beam directional antennas (MBA) more feasible [2-14]. Since the use of directional antennas divides the space around them in different beams, a multi-beam antenna with *m* beams can allow multiple (up to *m*) concurrent packet transmissions (CPT) or receptions (CPR) by a node using the same spatially overlapped channel, thus improving the throughput by up to *m* times as compared to the single beam directional antenna [1-4]. However, the network protocols which consider the use of omni-directional or single-beam directional communication cannot support the use of multi-beam directional communication. Therefore, new protocols (including the medium access control (MAC) and routing protocols) are needed for multi-beam directional communication.

In this paper, we describe our modeling and simulation framework to design an IEEE 802.11 DCF-based, multi-beam MAC scheme for multi-hop wireless networks on the Riverbed Modeler Wireless Suite [15]. The proposed work is important because none of the multi-beam MAC schemes available in literature (i.e., [4-8,10,13,16-27]) provide

This work was supported by the U.S. Department of Defense under Grant No. FA8750-14-1-0075. DISTRIBUTION STATEMENT A: Approved for Public Release; distribution unlimited 88ABW-2017-6519 28 Dec 2017.

their implementation details. Further, the network simulation tools, such as Riverbed Modeler, NS3, and QualNet, do not provide any multi-beam directional module. We have selected the Riverbed Modeler (formerly known as OPNET) in this paper because it is the most widely used commercial discrete event network simulator, which uses a well-defined physical layer and provides a flexible and accurate platform to evaluate the performance of variety of network protocols.

### 1.1 Related Work:

Limited information on the single-beam directional antenna pattern design in Riverbed Modeler was provided in [28]. An angular MAC (AN-MAC) scheme was proposed for Riverbed Modeler in [18-19], to reduce the deafness and collision problems in a directional ad-hoc network. Four antennas (each having a directional beam pattern of $90^0$ bandwidth like [28]) were connected to the PHY layer in AN-MAC and the setup was considered as a four-beam antenna. The CPT was achieved by transmitting the modified AN_RTS (angular RTS) and AN_CTS (angular CTS) control packets on different antennas simultaneously in [18-19]. However, this scheme cannot facilitate CPR. The AN-MAC scheme uses a DWTS (Directional Wait to Send) [29] like extension and transmits a packet, consisting of *dummy bits*, to silence the neighbor transmitter node(s). Despite having the CPT capability, AN-MAC can transmit only one data packet at a time which degrades spatial utilization. The modeling of the radio (i.e., transmitter and receiver port modules) to facilitate adaptive antennas in Riverbed Modeler was explained in [30]. Since it was designed for synchronous MAC protocol, it is not suitable for the more generalized IEEE 802.11 DCF CSMA/CA MAC, which is an asynchronous MAC protocol.

To the best of our knowledge, none of the existing works provides a complete framework to implement a multi-beam MAC scheme. Since this paper focuses on implementing a generalized multi-beam directional MAC scheme in Riverbed Modeler, we have selected HMAC scheme [8] as the basis of our implementation. Note that HMAC is the most widely cited and most comprehensive CSMA/CA based MAC protocol for multi-beam directional networks, and many multi-beam MAC schemes have used its beam synchronization, deafness, hidden terminal and head-of-line blocking mechanisms in their design (see [24-26] for recent examples). In addition to addressing the implementation challenges of HMAC scheme [8], we have also added new functionalities to (*a*) accommodate different retransmission scenarios, (*b*) support multi-hop transmission(s), and (*c*) address the issue of multipath.

### 1.2 Contributions:

The implementation of a basic multi-beam MAC protocol in Riverbed Modeler requires the development of a new multi-beam node model, and many other changes and additions in various source code modules in the Riverbed process model and pipeline stages for implementing the following:

i. Multi-beam antenna model to achieve the capability to transmit and receive multiple packets concurrently (i.e., CPT and CPR). It includes adding new directional beam patterns in the antenna pattern editor and linking them to the process model of the wireless_lan_mac module (a MAC layer module in the node model).

ii. Multiple sources (with independent traffic generation parameters) to generate data packets for different destinations. It requires changes in the network model and process model of the wlan_mac_intf module in the node model.

iii. Node-based contention window and collision handling mechanism at a multi-beam node for CPR and channel fairness. Modifications are made inside the wireless_lan_mac module.

iv. HNAV (hybrid network allocation vector) table in the wireless_lan_mac module to keep track of per beam information, such as potential neighbor node(s), NAV, and DoA (direction of arrival).

v. A mechanism to handle different retransmission scenarios on a bottleneck node (which can communicate on up to *m* beams simultaneously). Identifying the type of retransmission helps in achieving the maximum utilization of the CPT at a bottleneck node.

vi. A mechanism to support multi-hop transmission on a node. It requires changes in the wireless_lan_mac module.

vii. A new mechanism is added in the wireless_lan_mac module to resolve the issue of multipath, where the copies of a packet arrive at different antenna ports of a bottleneck node, due to the non-line-of-sight propagation.

The researchers can implement their directional MAC and routing protocols (single-beam as well as multi-beam) in Riverbed Modeler by using the modules and methodology discussed in this paper. In fact, a multipath routing scheme for the MBA equipped nodes, which used our proposed multi-beam MAC protocol framework, has appeared in [27].

### 1.3 Paper Organization:

This paper is organized in six sections. We discuss the overview and key features of a basic multi-beam MAC protocol in Section 2. In Section 3, the network architecture of Riverbed Modeler is described. The implementation of multi-beam MAC scheme, the challenges faced during the implementation, and the solution proposed for them are presented in Section 4. Simulation setup and results are discussed in Section 5, followed by conclusion in Section 6.

## 2. OVERVIEW OF THE MULTI-BEAM MAC PROTOCOL

The IEEE 802.11 DCF was designed for omni-directional communication, which limits spectral efficiency. A multi-beam node (known as the bottleneck node) can transmit or receive multiple packets concurrently on its different beams. However, all beams of a half-duplex bottleneck node can either be in transmission (CPT) or in reception (CPR) mode. The MBA is assumed to be a switched beam antenna, which can switch multiple beams simultaneously. It is a wide-azimuth antenna which covers the $360^0$ through its *m* non-overlapping beams such that each beam covers $360^0/m$ angle. The total transmit power is equally divided among these *m* beams, and each beam performs its own independent channel sensing directionally. The key features of a multi-beam MAC protocol are discussed below.

In order to support multi-beam communication, the MAC layer functionality has been redefined as described below [8]. To achieve CPT, all the active beams of a bottleneck node must be synchronized. A bottleneck node also requires each intended transmitter node on its different beams to transmit simultaneously, which ensures the CPR if propagation delay is neglected. Since the node deafness and hidden terminal problems increase significantly in directional communication, a bottleneck node should transmit control packets on those beams which contain a potential transmitter for it. In the IEEE 802.11 DCF MAC, the contention window (CW) of a node decreases to the minimum after each successful transmission, whereas it is doubled for each unsuccessful transmission until the maximum retry limit is reached. During concurrent communications on its *m* beams, a bottleneck node may experience some unsuccessful transmissions. To address this issue, the node-based backoff can be used, in which a bottleneck node maintains a single CW for its all beams. If at least one beam achieves successful transmission, the bottleneck node's CW is set to minimum. Further, a node extracts the packet from its data buffer queue and transmits it through its antenna. To avoid the HOL (head of line) packet blocking problem [1], a bottleneck node must maintain multiple data buffer queues at its MAC layer so that it can concurrently transmit packets on its multiple beams.

### 2.1 Key Features

#### *2.1.1* ***Removal of Random Backoff***

To reduce the probability of more than one node accessing the same channel simultaneously, each node in IEEE 802.11 waits a random backoff duration after sensing the channel free for the DIFS (distributed inter frame space) duration. However, all transmitter nodes should transmit simultaneously to achieve CPR at the bottleneck node. Hence, random backoff is removed so that the transmitter nodes can be locally synchronized [8].

#### *2.1.2* ***Scheduling Packets and the HNAV Table***

A receiver node is considered as deaf to a transmitter if its beam is not directed towards that transmitter node. Being deaf towards the incoming transmission results in increased backoff delay at the affected transmitter node. Once the packet's transmission count reaches the maximum retry limit, the MAC layer drops it and requests the network layer to initiate a fresh route discovery. A bottleneck node reduces the deafness of its neighboring node by transmitting the scheduling (SCH) packets on those beams which contain potential transmitters for it. Each bottleneck node maintains the HNAV (hybrid network allocation vector) table for this purpose, which contains the per beam information, such as neighbor node's address, NAV, and two Boolean variables (*isValidRTSReceived* and *isInvalidCTSReceived*).

If a beam receives the valid RTS (or invalid CTS), it sets its *isValidRTSReceived* (or *isInvalidCTSReceived*) variable to true. These Boolean variables help a node in determining if it has a potential transmitter in its neighborhood. If it does have a potential transmitter, the bottleneck node sends an SCH/RTS or SCH/CTS control packet, whose structure is similar to the corresponding RTS or CTS packet, and updates the NAV of that beam (for beam synchronization purpose). The neighbor node(s), upon receiving the SCH packet, updates its NAV for the duration until it needs to defer its transmission.

#### *2.1.3* ***Mitigating Node Starvation and Improving Network Fairness***

Without the random backoff, all locally synchronized nodes start their transmissions concurrently. If these transmitter nodes have high data generation rates, they can prevent the receiver bottleneck node from accessing the channel, which leads to its starvation. To address this issue, a role priority switching (i.e., the hot-potato mechanism) is used in order to minimize the queuing delays by having successive cycles of CPT and CPR on a node [8]. It gives time to the receiver node to start its own transmission. However, it is possible that the receiver bottleneck node has other transmitter nodes in its neighborhood. As these nodes (the bottleneck receiver node and its neighboring transmitter nodes) do not have active NAV, they can start their transmissions simultaneously which would make them deaf to each other. To address this deafness issue, the jump backoff mechanism is used, in which the receiver node adds AIFS (additional inter frame space) time in the duration field of SCH/CTS packet [8]. It forces these neighboring transmitter nodes to wait for a longer duration. Meanwhile, the receiver bottleneck node starts its own transmission if the channel is unoccupied, followed by the role priority switching so that its neighboring transmitter nodes can access the channel. Thus, the role priority switching and jump backoff mechanisms, together, improve network fairness.

#### *2.1.4* ***Protocol Operation***

Fig. 1 shows an example of the network topology. Here X, Y, and Z are bottleneck nodes (i.e., equipped with MBA). Node X has packets for nodes Y and Z. Nodes 1 and 2 also have packets for node Y. Nodes 3 and 4 are in the neighborhood of node Z and node X, respectively. The basic operation of a multi-beam protocol was explained in Fig. 9 in [8]. After sensing the channel free for the DIFS duration, source X transmits RTS packets to nodes Y and Z, and SCH/RTS packet towards node 4. Upon receiving RTS, nodes Y and Z wait for SIFS (short inter frame space) duration before replying with CTS. Meanwhile, node 4 updates its NAV. Being an intermediate node, node Y needs to forward this packet to node 1. Hence, it adds the AIFS duration (as part of the jump backoff mechanism) and sends SCH/CTS packet to nodes 1 and 2. Since node Z does not have any packet in its data buffer queue, it does not add the AIFS duration in its SCH packet. Hence the updated NAV of node 3 (neighborhood of Z) is less than node Y's neighbor node's NAV. Once nodes Y and Z successfully receive data packets, they send ACK packets. Node X, upon receiving ACK packets, performs role priority switching, which leaves the channel free for the intermediate node Y. So, node Y transmits RTS towards node 1.

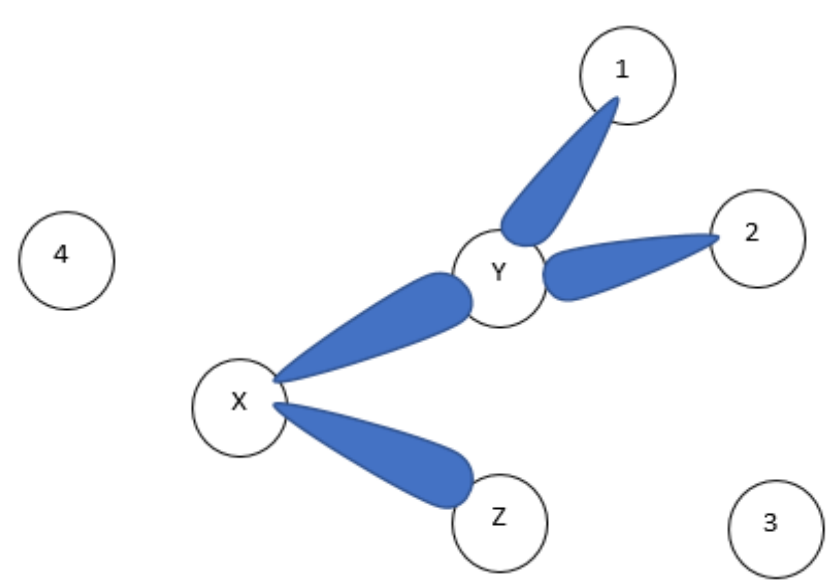

Fig. 1. An example of the network topology.

## 3. NETWORK SIMULATION ARCHITECTURE IN RIVERBED MODELER

Riverbed Modeler (formerly known as OPNET) is a discrete event-based network simulator [15]. Events are specific tasks which occur at a certain time. When an event occurs, simulator updates its simulation time. Riverbed Modeler decomposes the complex layered protocols, such as IEEE 802.11, into more manageable parts with well-defined interfaces to adjacent layers. Riverbed Modeler uses a hierarchical architecture consisting of the network model, the node model, and the process model with their corresponding editors, which are discussed below.

**3.1** The **network model** is built on the graphical user interface (GUI) provided by the project editor window. This is where network topologies are designed, nodes are configured, trajectories are assigned, random noise is imported, channel behavior is shaped, and local or global statistics are observed. A hybrid environment of fixed and mobile nodes can be created by customizing the node's *edit attributes*. A topology can be a mixture of point-to-point links, bus links or wireless radio links. User-defined network variables can be introduced through extended attributes as shown in Fig. 2.

**3.2** The **node model** (Fig. 3) provides a high-level overview of data packet flow within different layers. The source, sink, MAC interface (wlan_mac_intf), MAC (wireless_lan_mac), and transmitter (radio_tx1, etc.) and receiver (radio_rx1, etc.) pair are few of the basic modules a node model must have. All the modules are connected either through wires or wireless links. The wires are of two types, stream wire and statistic wire. The stream wires are used to handover large information (such as the data packets) from one module to another, whereas the statistic wires are used to transfer relatively small information (such as interrupt value). The stream wires are represented by solid lines (in red and blue colors), and the statistic wires by dotted lines (brown color) in Fig. 3. The handover can be either *quiet*, *scheduled*, or *forced*. In the *scheduled* mode, an interrupt is triggered and the information is passed only when the current event is finished. In the *quiet* mode, information is delivered without triggering any interrupt. In the *forced* mode, the information is forcefully delivered by suspending the current event.

Physical (PHY) layer functionality is encapsulated inside the *edit attributes* of the transmitter and receiver port pair in the form of 14 pipeline stages (Fig. 4). Riverbed Modeler provides the flexibility to include modified or newly developed stages. For each transmission, stages 1 (transmission delay) to 5 (propagation delay) are invoked. Stages 6 (receiver antenna gain) to 13 (error correction) are invoked for each reception. Stage 0 (receiver group) is optional and is invoked only once, at the start of the simulation. The main objectives of these stages are: to establish a functional link between the transmitter and receiver nodes, to transmit and receive the packets, to calculate different types of delay (transmission delay, propagation delay, etc.), to calculate signal to noise ratio (SNR), to correct errors (if possible), and finally to forward it to the MAC. The stage 3 (channel match) assigns either of these three statuses - *valid*, *noise*, and *ignored*, to the packet before its transmission. Any pipeline stage can change the status of the packet. Only those packets reach the MAC layer which manage to retain their *valid* status throughout the PHY layer processing.

The antenna beam patterns with varying directivity can be designed and realized in 2D and 3D through the antenna port's gain table. Riverbed Modeler provides support to export or import gain table to (or from) any ASCII file. For a more practical setting, side and minor lobes should be taken into consideration, by normalizing the gain table. Fig. 5 shows one such gain table, where the elevation (polar) and azimuth planes have a half power beam width (HPBW) of $10^0$ each.

**3.3** Each module of a node model leads to a **process model**, which is a combination of the forced and unforced finite state machines (Fig. 6). Each of these states contains the Enter and Exit executives, written in Proto-C. The process model editor provides the workspace to write these scripts. Unlike the forced state (green color), the unforced states (red color) need the required interrupt to be triggered before jumping to the Exit executive from its Enter executive. The process models must be compiled successfully in order to run the network simulation.

There are six different code blocks inside every process model: state variables (SV), temporary variables (TV), header block (HB), diagnostic block (DB), termination block (TB), and function block (FB). The global variables are declared in SV. The variables stored in TV are destroyed once a state is switched. The HB contains all the header files, constant variables, enumerated types, definitions for data structures, macros, and function prototypes, etc. The DB is used in diagnosing any situation. Writing a TB script makes sure that the system will get its allotted memory space back before destroying a process. FB is the main block where most of the functions get their definitions. The important functions in FB are: wlan_mac_sv_init(), wlan_higher_layer_data_arrival(), wlan_hl_packet_drop(), wlan_hlpk_enqueue(), wlan_frame_transmit(), wlan_prepare_frame_to_send(), wlan_interrupts_process(), wlan_physical_layer_data_arrival(), wlan_tuple_find(), wlan_data_process(), wlan_schedule_deference(), wlan_frame_discard(), wlan_mac_rcv_channel_status_update(), wlan_sta_addr_register(), and wlan_reset_sv(), etc. Their details can be found in Riverbed Modeler documentation [15].

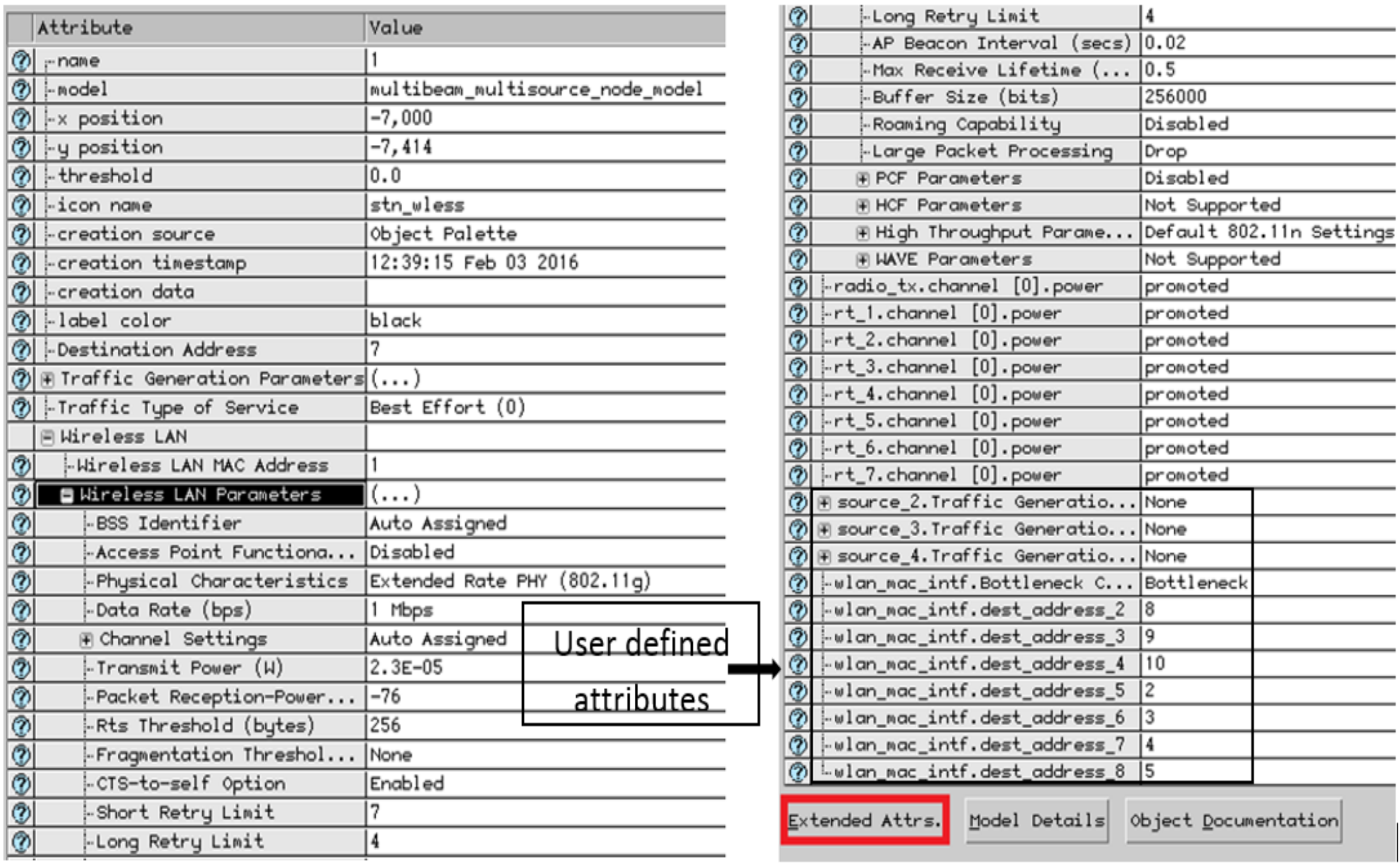

| Attribute | Value |
|---|---|
| name | 1 |
| model | multibeam_multisource_node_model |
| x position | -7,000 |
| y position | -7,414 |
| threshold | 0.0 |
| icon name | stn_wless |
| creation source | Object Palette |
| creation timestamp | 12:39:15 Feb 03 2016 |
| creation data | |
| label color | black |
| Destination Address | 7 |
| Traffic Generation Parameters | (...) |
| Traffic Type of Service | Best Effort (0) |
| Wireless LAN | |
| Wireless LAN MAC Address | 1 |
| Wireless LAN Parameters | (...) |
| BSS Identifier | Auto Assigned |
| Access Point Functiona... | Disabled |
| Physical Characteristics | Extended Rate PHY (802.11g) |
| Data Rate (bps) | 1 Mbps |
| Channel Settings | Auto Assigned |
| Transmit Power (W) | 2.3E-05 |
| Packet Reception-Power... | -76 |
| Rts Threshold (bytes) | 256 |
| Fragmentation Threshol... | None |
| CTS-to-self Option | Enabled |
| Short Retry Limit | 7 |
| Long Retry Limit | 4 |
| Long Retry Limit | 4 |
| AP Beacon Interval (secs) | 0.02 |
| Max Receive Lifetime (... | 0.5 |
| Buffer Size (bits) | 256000 |
| Roaming Capability | Disabled |
| Large Packet Processing | Drop |
| PCF Parameters | Disabled |
| HCF Parameters | Not Supported |
| High Throughput Parame... | Default 802.11n Settings |
| WAVE Parameters | Not Supported |
| radio_tx.channel [0].power | promoted |
| rt_1.channel [0].power | promoted |
| rt_2.channel [0].power | promoted |
| rt_3.channel [0].power | promoted |
| rt_4.channel [0].power | promoted |
| rt_5.channel [0].power | promoted |
| rt_6.channel [0].power | promoted |
| rt_7.channel [0].power | promoted |
| source_2.Traffic Generatio... | None |
| source_3.Traffic Generatio... | None |
| source_4.Traffic Generatio... | None |
| wlan_mac_intf.Bottleneck C... | Bottleneck |
| wlan_mac_intf.dest_address_2 | 8 |
| wlan_mac_intf.dest_address_3 | 9 |
| wlan_mac_intf.dest_address_4 | 10 |
| wlan_mac_intf.dest_address_5 | 2 |
| wlan_mac_intf.dest_address_6 | 3 |
| wlan_mac_intf.dest_address_7 | 4 |
| wlan_mac_intf.dest_address_8 | 5 |

Fig. 2. Node's *edit attributes* with user-defined variables [15].

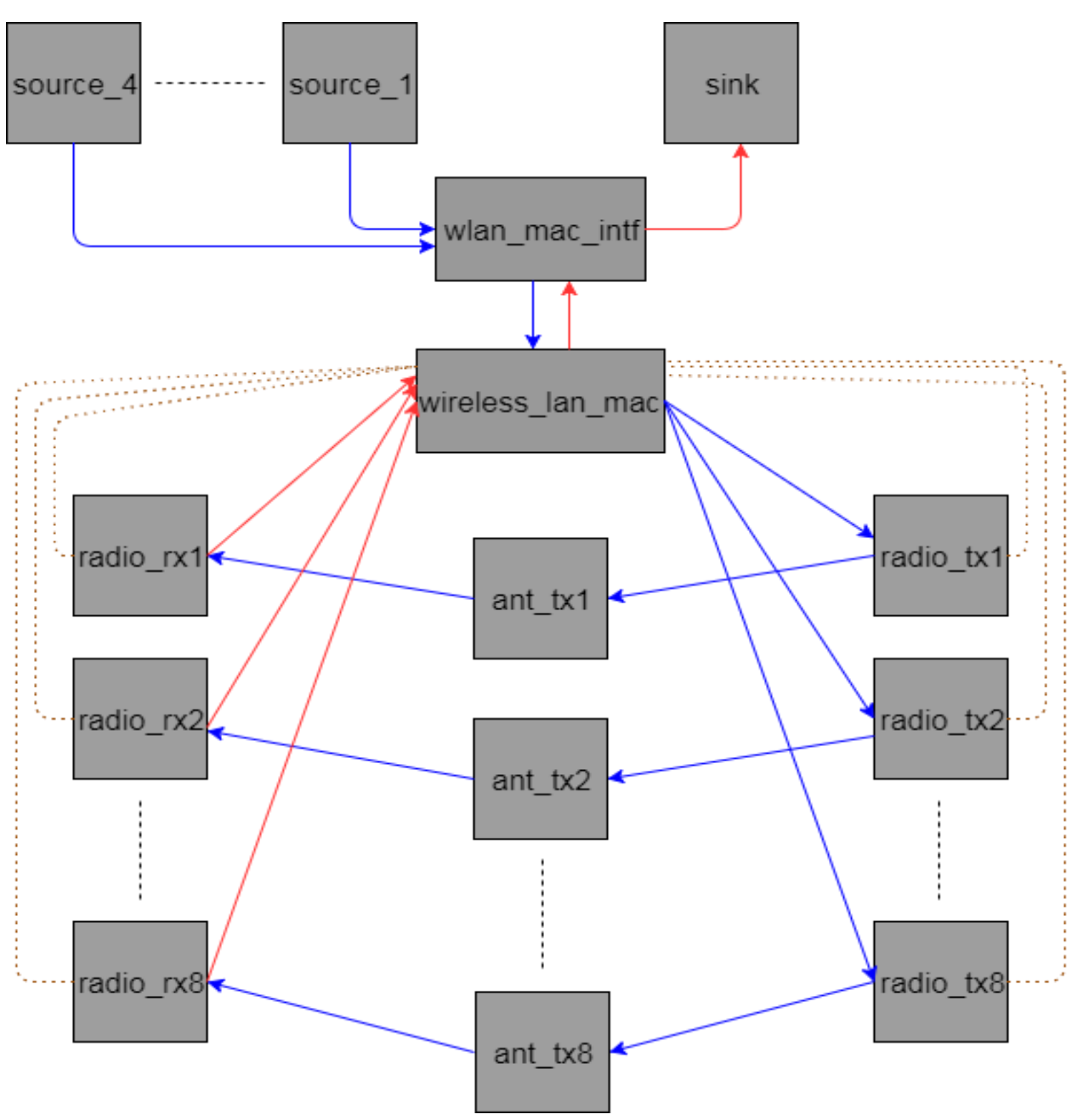

Fig. 3. The multi-beam node model.

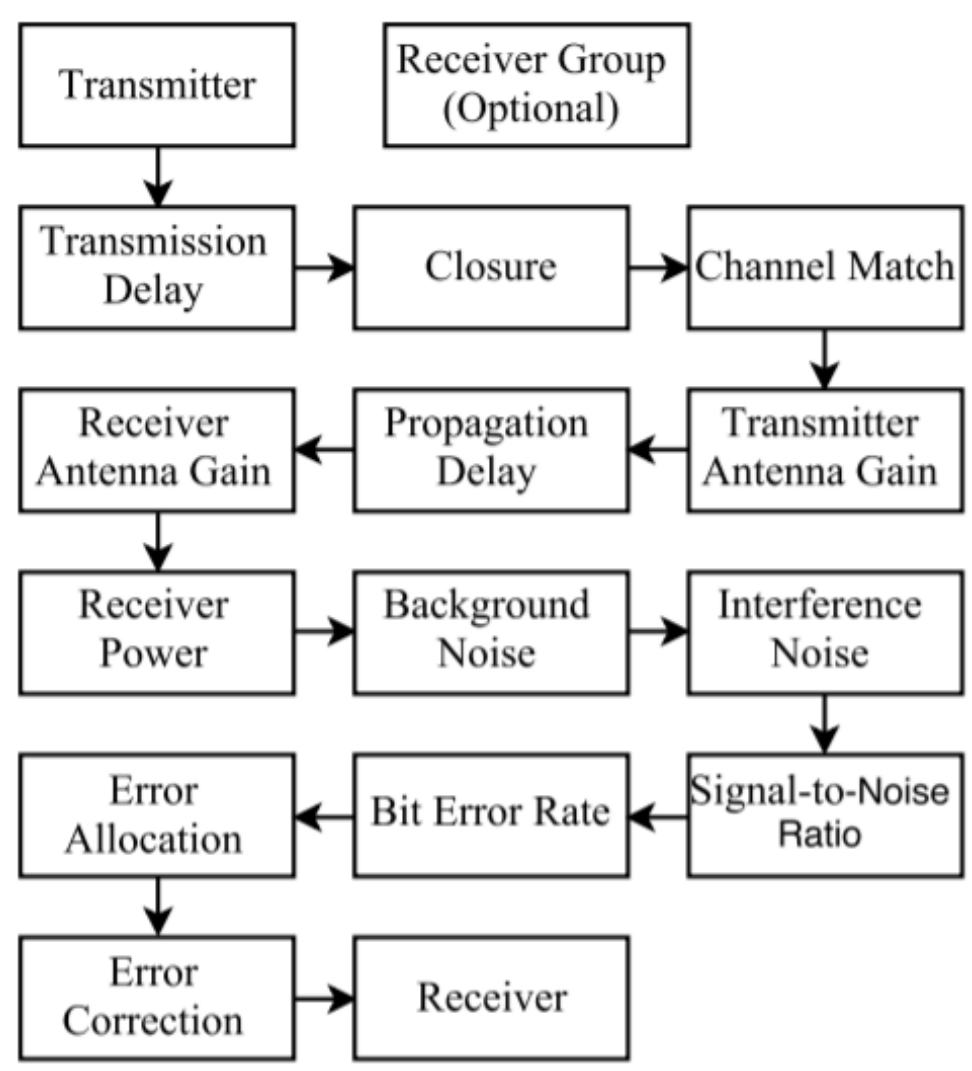

Fig. 4. The pipeline stages in Riverbed Modeler (Source [15]).

Polar Plane Section | Polar\Azimuth Gain Table

Polar (rows) and Azimuth (columns) values in degrees, gain cell values in dB.

| | 0 | 5 | 10 | 15 | 20 | 25 | 30 | 35 | 40 | 45 | 50 |
|---|---|---|---|---|---|---|---|---|---|---|---|
| 0 | -0.087 | -0.087 | -0.087 | -0.087 | -0.087 | -0.087 | -0.087 | -0.087 | -0.087 | -0.087 | -0.087 |
| 5 | -0.087 | -0.087 | -0.087 | -0.087 | -0.087 | -0.087 | -0.087 | -0.087 | -0.087 | -0.087 | -0.087 |
| 10 | -0.087 | -0.087 | -0.087 | -0.087 | -0.087 | -0.087 | -0.087 | -0.087 | -0.087 | -0.087 | -0.087 |
| 15 | -0.087 | -0.087 | -0.087 | -0.087 | -0.087 | -0.087 | -0.087 | -0.087 | -0.087 | -0.087 | -0.087 |
| 20 | -0.087 | -0.087 | -0.087 | -0.087 | -0.087 | -0.087 | -0.087 | -0.087 | -0.087 | -0.087 | -0.087 |
| 25 | -0.087 | -0.087 | -0.087 | -0.087 | -0.087 | -0.087 | -0.087 | -0.087 | -0.087 | -0.087 | -0.087 |
| 30 | -0.087 | -0.087 | -0.087 | -0.087 | -0.087 | -0.087 | -0.087 | -0.087 | -0.087 | -0.087 | -0.087 |
| 35 | -0.087 | -0.087 | -0.087 | -0.087 | -0.087 | -0.087 | -0.087 | -0.087 | -0.087 | -0.087 | -0.087 |
| 40 | -0.087 | -0.087 | -0.087 | -0.087 | -0.087 | -0.087 | -0.087 | -0.087 | -0.087 | -0.087 | -0.087 |
| 45 | -0.087 | -0.087 | -0.087 | -0.087 | -0.087 | -0.087 | -0.087 | -0.087 | -0.087 | -0.087 | -0.087 |
| 50 | -0.087 | -0.087 | -0.087 | -0.087 | -0.087 | -0.087 | -0.087 | -0.087 | -0.087 | -0.087 | -0.087 |
| 55 | -0.087 | -0.087 | -0.087 | -0.087 | -0.087 | -0.087 | -0.087 | -0.087 | -0.087 | -0.087 | -0.087 |
| 60 | -0.087 | -0.087 | -0.087 | -0.087 | -0.087 | -0.087 | -0.087 | -0.087 | -0.087 | -0.087 | -0.087 |
| 65 | -0.087 | -0.087 | -0.087 | -0.087 | -0.087 | -0.087 | -0.087 | -0.087 | -0.087 | -0.087 | -0.087 |
| 70 | -0.087 | -0.087 | -0.087 | -0.087 | -0.087 | -0.087 | -0.087 | -0.087 | -0.087 | -0.087 | -0.087 |
| 75 | -0.087 | -0.087 | -0.087 | -0.087 | -0.087 | -0.087 | -0.087 | -0.087 | -0.087 | -0.087 | -0.087 |
| 80 | -0.087 | -0.087 | -0.087 | -0.087 | -0.087 | -0.087 | -0.087 | -0.087 | -0.087 | -0.087 | -0.087 |
| 85 | -0.087 | -0.087 | -0.087 | -0.087 | -0.087 | -0.087 | -0.087 | -0.087 | 25.023 | 25.023 | 25.023 |
| 90 | -0.087 | -0.087 | -0.087 | -0.087 | -0.087 | -0.087 | -0.087 | -0.087 | 25.023 | 25.023 | 25.023 |
| 95 | -0.087 | -0.087 | -0.087 | -0.087 | -0.087 | -0.087 | -0.087 | -0.087 | 25.023 | 25.023 | 25.023 |
| 100 | -0.087 | -0.087 | -0.087 | -0.087 | -0.087 | -0.087 | -0.087 | -0.087 | -0.087 | -0.087 | -0.087 |

Fig. 5. A portion of the antenna gain table in Riverbed Modeler [15].

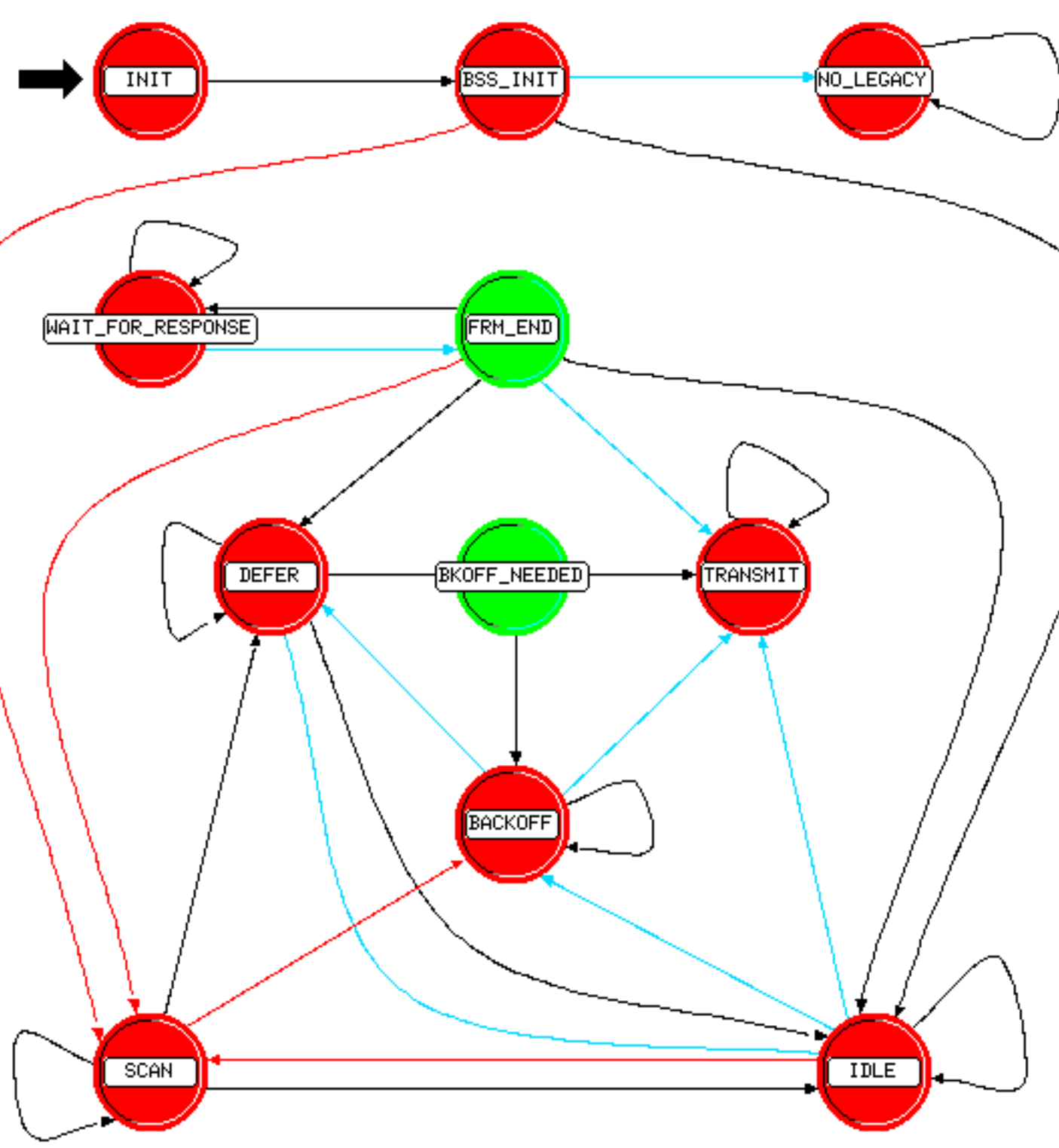

Fig. 6. The process model in Riverbed Modeler (adapted from [15]).

## 4. IMPLEMENTATION OF MULTI-BEAM MAC PROTOCOL IN RIVERBED MODELER

Like most network simulators, Riverbed Modeler has the modules to support the omni-directional IEEE 802.11 MAC protocol. As discussed in Section 1.1, very little work on the implementation of a directional MAC protocol in Riverbed Modeler is available in the literature. Riverbed Modeler encapsulates the functionality of the IEEE 802.11 MAC layer in the wireless_lan_mac module. Inside it, we can select child processes, such as wlan_mac (for DCF mechanism), and wlan_mac_hcf (for hybrid coordination mode mechanism). Various challenges that we faced during the implementation of the multi-beam MAC protocol in Riverbed Modeler and the proposed solutions to overcome them are described below. The modifications made in the antenna module, process model of the wireless_lan_mac module, source module, and wlan_mac_intf module are also explained below.

### 4.1. Multi-Beam Antenna Model Design

The multi-beam antenna module is not available in Riverbed Modeler. Like [28], we are using multiple antennas to create our multi-beam antenna module. In our implementation, each beam corresponds to a specific antenna, which has its own customized and non-interfering gain pattern (as shown in Fig. 5 for one beam). The node model in Fig. 3 shows eight Tx-Rx port pairs, associated with different antennas (ant_tx1 to ant_tx8). These Tx-Rx port pairs are attached to different indices of the wireless_lan_mac module (located in the node model) through statistic and stream wires. This setup allows the MAC layer, which resides inside the wireless_lan_mac module, to interact with a specific antenna port independently. Because of it, the MAC layer forwards packets to different antenna ports, through corresponding Tx ports to achieve CPT. Similarly, when any beam (i.e., the antenna) receives a packet, its corresponding Rx port passes the packet through its pipeline stages and then forwards to the MAC layer.

#### *4.1.1* ***Antenna Gain Calculation***

We use a generalized multi-beam antenna with planner array elements and eight non-overlapping beams, where each beam has $10^0$ HPBW in both elevation ($\theta$) and azimuth ($\emptyset$) planes. The antenna gain mainly depends on the HPBWs and antenna array structure. For planner arrays, the approximation between directivity (*D*) and HPBWs for directional patterns, which have one main lobe and negligible side lobes, is given by [31],

$$D_0 = \frac{32400}{(HPBW_{azimuth(deg)} * HPBW_{elevation(deg)})} \tag{1}$$

The antenna gain (G$(\theta, \emptyset)$) can be computed from antenna directivity and radiation efficiency ($e_{cd}$) as [31],

$$G(\theta,\emptyset) = e_cd * D(\theta,\emptyset) \tag{2}$$

If $e_{cd}$ is 100%, the maximum directivity ($D_0$) of the antenna becomes equal to its maximum gain ($G_0$). From Eq. (1), for $10^0$ HPBW, $D_0$ is 324 (non-dB) or 25.11dB. After normalizing the gain, the main and side lobe gains are 25.023dB and -0.087dB, respectively (see Fig. 5). The relationship between transmission power ($P_t$) and coverage range (*R*) can be obtained from Friis transmission equation as [31],

$$\frac{P_r}{P_t} = G_r * G_t * (\frac{\lambda}{4\pi R})^2 \tag{3}$$

Here $G_r$ and $G_t$ are the gains of the Rx and Tx antennas, respectively. $P_r$ is the receiver power and $\lambda$ is the wavelength. The values of transmission power and packet reception-power threshold can be modified inside the node's *edit attributes* at the network model. We feed the same transmission power to each Tx port of the multi-beam antenna. Therefore, the directional coverage range of the bottleneck and non-bottleneck nodes are the same.

### 4.2 Enabling CPR Functionality at PHY Layer

Since the IEEE 802.11 MAC was designed for omni-directional communication, it does not support CPR. We use a global variable, called *rx_end_time*, in Riverbed Modeler to enable the CPR. Every time the Rx port starts receiving a packet, the Receiver Power stage (Stage 8 of the pipeline stages in Fig. 4) updates the *rx_end_time* with the time by which the reception of this packet will be finished. Although each Rx port executes its pipeline stages separately, these stages refer to the common global variables, such as the *rx_end_time*. If a beam receives a packet, it updates the *rx_end_time* (as a result, *rx_end_time* > current time). When another beam of that bottleneck receiver node receives a packet simultaneously, the beam would consider its packet as noise because the *rx_end_time* > current time. Furthermore, the Interference Noise stage (Stage 10 of the pipeline stages) updates the status of both the received packets to *noise* because a node is not supposed to receive multiple packets concurrently in Riverbed Modeler. As a result, none of the packets reaches the MAC layer.

To overcome this issue, we ignore the value of *rx_end_time* at the Receiver Power stage (Stage 8 of the pipeline stages) to allow multiple beams to receive packets concurrently. However, this leads to another problem, i.e., when an Rx port receives multiple packets, all of them get accepted by the Receiver Power stage. To address this issue, we compare the received power of all the packets that arrive concurrently on the same Rx port, at the Interference Noise stage (Stage 10 of the pipeline stages). The packet with maximum power retains its status, whereas other packets are *ignored*. Finally, the Error Correction stage (Stage 13 of the pipeline stages) forwards packet(s) having *valid* status to the MAC layer while discarding the remaining packets.

### 4.3 Multiple Sources and Buffer Queues

To address the HOL packet blocking problem [1], we created four data buffer queues at the MAC. In the multi-beam node model (Fig. 3), four sources are used which can have independent traffic generation parameters (Fig. 2). In Riverbed Modeler, the wlan_mac_intf module (located inside the node model) receives these packets from the sources and passes them (one packet at a time) to the MAC along with an ICI (information control interface) packet, called *wlan_mac_req_iciptr*, as shown in Fig. 7. The ICI packet is much smaller than the data packet and is used to pass information, such as the protocol type and destination address. As the MAC has multiple buffer queues, the packet (arriving from the network layer) needs to be sorted so that it can be stored in the appropriate queue. To achieve this, we have added a new field, called *queue_index*, in the ICI packet, which also represents the *source_index*.

Since Riverbed Modeler does not allow the ICI packet to be copied [32], the ICI pointer is reinitialized when more than one data packets arrive at the wlan_mac_intf process model concurrently (as shown in Fig. 7). Although the MAC receives all these data packets, they use the reinitialized ICI packet. To avoid this implementation issue, the traffic generation parameters of different sources should be selected such that wlan_mac_intf does not receive multiple data packets simultaneously (see Section 5.1 for more details).

When the MAC layer receives a data packet from the PHY layer, its final destination address is checked. If the receiver bottleneck node is not the final destination address, the MAC stores the received packet in one of its queues (selection of the *queue_index* for an intermediate receiver bottleneck node is discussed in Section 4.10). Whereas, if this node is the final destination address, the MAC forwards the received packet to the sink module through the wlan_mac_intf module (see Fig. 3).

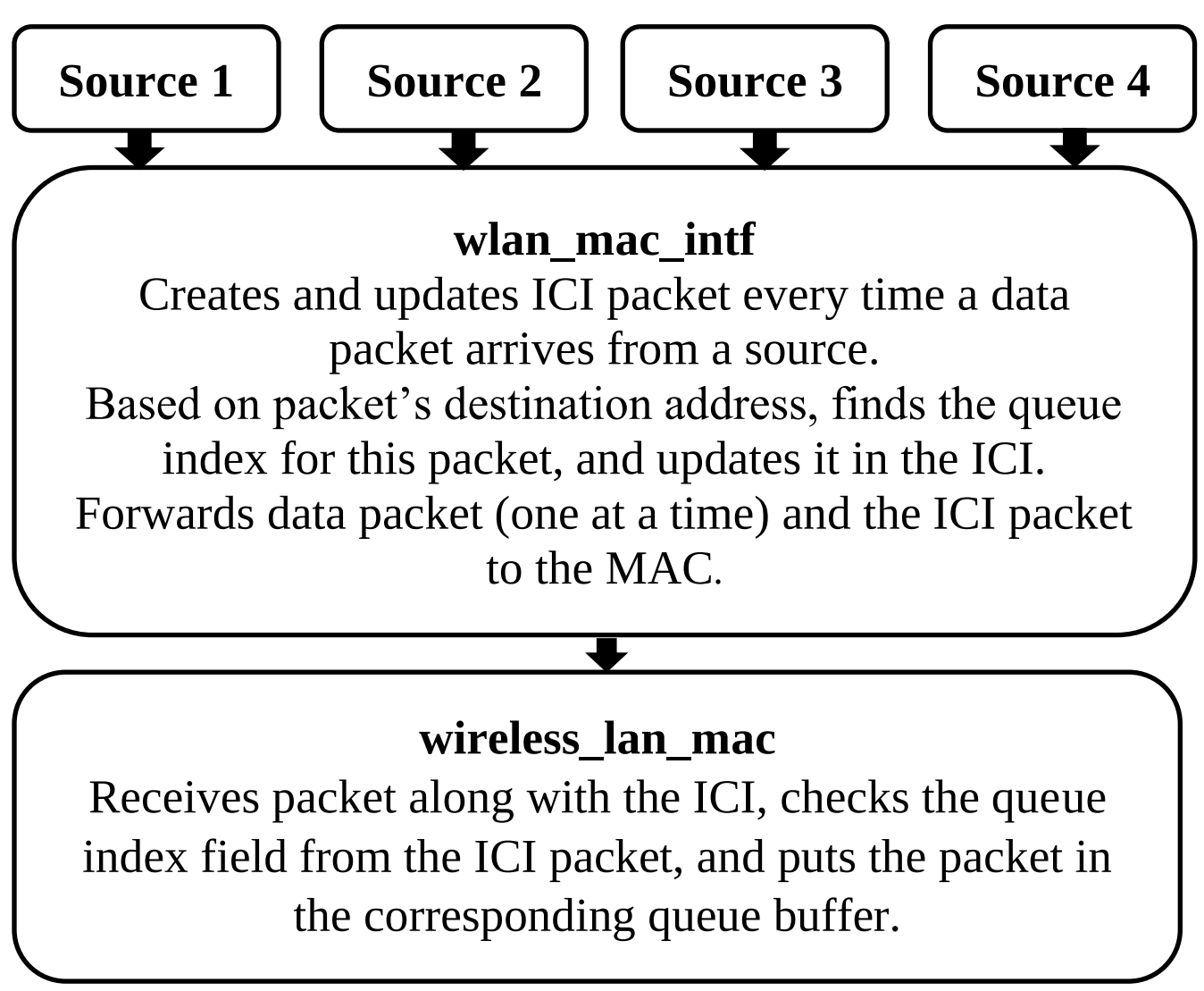

Fig. 7. Packet flow from the source(s) to MAC layer.

### 4.4 Enabling CPT Functionality at MAC Layer

Different beams of a multi-beam node perform their functionality independently. Therefore, the MAC layer must synchronize them to achieve the CPT. For this purpose, we introduce a new Boolean flag, called *to_Send*, in the HNAV table (located inside the function block of the wireless_lan_mac module). A beam sets this flag to true when it has a non-SCH packet to transmit. The beams, having the *to_Send* flag true, forward their packets to the PHY layer concurrently and achieve CPT.

### 4.5 Use of Contention Window

After each successful transmission, the CW of a node resets to the minimum ($CW_{min}$). For IEEE 802.11 standard, the values of $CW_{min}$ and $CW_{max}$ are 15 and 1023, respectively. A node randomly selects its CW value from the uniformly distributed interval [0, (CW+1)] and then waits for those many slots before the transmission. In contrast, a constant backoff is used in order to obtain the locally synchronized nodes, in order to achieve CPR [8].

We have modeled the above node-based backoff approach in the bkoff_needed state, which is located in the process model of the wireless_lan_mac module in Riverbed Modeler (see Fig. 6). When a node achieves successful transmission, it resets its CW to $CW_{min}$, and waits for the constant backoff before transmission. Whereas for each unsuccessful transmission, its CW is doubled. The pseudo code is given below. Riverbed Modeler uses wlan_flags structure to track the current state of the node, and OPC_TRUE as Boolean true.

```
if(wlan_flags→backoff_flag == OPC_TRUE || wlan_flags→perform_cw == OPC_TRUE) {
        if (last transmission was successful || wlan_flags→perform_cw == OPC_TRUE)
                max_backoff = cw_min;    // max_backoff temporarily holds the value of backoff slots
        else // increase the back-off window for retransmission
                max_backoff = max_backoff  * 2 + 1;
        if (wlan_flags→perform_cw == OPC_TRUE)
                backoff_slots = max_backoff + 1;
        else
                //backoff_slots = floor(op_dist_uniform(max_backoff + 1)); // used in IEEE 802.11 MAC
                backoff_slots = max_backoff + 1;   // modified for the multi-beam MAC protocol
}
```

### 4.6 Handling Collisions at the Receiver Bottleneck Node

In a hidden terminal situation, more than one packet can arrive asynchronously at the common receiver node. If the receiver node is deferring for the SIFS duration and a new packet arrives, its PHY layer does not change the *valid* status of the packet. As a result, the MAC layer receives multiple packets and suspends its functionality. To address this issue, the MAC layer flags the received frames as invalid and waits for EIFS (extended inter frame space) duration. Such hidden terminal situation can arise where the MAC layer receives more than one packet from the same Rx port.

In Riverbed Modeler, when the PHY layer senses a busy channel, the corresponding Rx port, through its connecting statistic wire, schedules an interrupt inside the function block of the wlan_lan_mac module. This triggers a call to the wlan_mac_rcv_channel_status_update() function to update node's busy state. The function uses a Boolean variable, called *receiver_busy*, for this purpose. The pseudo code for flagging the collision at the MAC layer (for the IEEE 802.11 MAC) is given below.

```
each time the channel is busy and the received power > rx_power_threshold {
        if (!wlan_flags→collision && wlan_flags→receiver_busy == OPC_TRUE)
                wlan_flags→collision = OPC_TRUE;
                Defer for EIFS duration
        if (!wlan_flags→receiver_busy)
                wlan_flags→receiver_busy = OPC_TRUE;
}
```

For an eight-beam antenna, we have included eight *receiver_busy* flags (i.e. receiver_busy1 to receiver_busy8) where each flag corresponds to a specific Rx port. The collision variable is flagged when the *receiver_busy* flag of its Rx port is already set but that Rx port again receives a packet. Since any Rx port can flag a collision, we introduced a new global Boolean variable, called *update_collision*, inside the function block of the wireless_lan_mac module. The pseudo code for flagging the collision is described below.

```
each time the channel is busy and the received power > rx_power_threshold {
        if (!wlan_flags→receiver_busy1)
                wlan_flags→receiver_busy1 = OPC_TRUE;
        …….
        else if(!wlan_flags→receiver_busy8)
                wlan_flags→receiver_busy8 = OPC_TRUE;
        else    // A Rx port has received more than one packet
                update_collision = OPC_TRUE;
        if (!wlan_flags→collision && update_collision)
                wlan_flags→collision = OPC_TRUE;
                update_collision = OPC_FALSE;   //Riverbed Modeler uses OPC_FALSE as Boolean false
                Defer for EIFS duration
}
```

Furthermore, the transition from one state to another state in the process model depends on the conditions of the macros defined in the header block. With the introduction of multiple *receiver_busy* flags, we have modified the definitions of some of these transitional macros: including medium_is_idle (transfers from idle state to either transmit or defer state), deference_off (transfers from defer to backoff_needed state), backoff_completed (used in perform_transmit macro), cw_completed (used in perform_transmit macro), receiver_busy_low (used in function block of the process model), and back_to_defer (transfers from backoff to defer state), etc.

### 4.7 Arrival of Undesired Packets

In Fig. 1, consider the situation where node 4 has a packet for node X, and node Y has packets for nodes 1 and 2. They both find the channel free and send RTS packets. If the bottleneck node X's beam, which is directed towards node Y, has *isValidRTSReceived* variable set to true, it sends SCH/CTS to node Y and CTS to node 4. Meanwhile, nodes 1 and 2 also send CTS to node Y.

If node X completes its SIFS wait before nodes 1 and 2, its SCH/CTS packet can reach node Y before other packets. In IEEE 802.11 MAC, each node waits for the frame timeout duration to get a response from its receiver. When a node's antenna starts receiving a packet, the MAC cancels the frame timeout interrupt. However, the cancellation of this interrupt, in this case, leads to an asynchronous packet arrival scenario.

Furthermore, the arrival of the undesired packet is problematic on those beams which are not expecting any packets. Therefore, before canceling the frame timeout interrupt, we compare the beam index (on which the packet arrives) with those beam indexes on which the node is expecting a response. Hence, here, node Y does not allow the SCH/CTS packet from node X to cancel the frame timeout interrupt and therefore achieves CPR.

### 4.8 Development of Generalized Node and Process Models

In our implementation, both bottleneck and non-bottleneck nodes share the same node and process models. In the node model, a new attribute, called *Bottleneck Capability*, is introduced which can be selected either as default (which means non-bottleneck) or bottleneck at the network model (e.g., the bottleneck capability is selected in the user-defined *edit attributes* in Fig. 2).

In Riverbed Modeler, the MAC layer uses the wlan_physical_layer_data_arrival() function (located inside the wireless_lan_mac module) to process the frames received from the lower layer. This function is not designed to process multiple packets. Therefore, a new function, called wlan_physical_layer_multiple_data_arrival(), is introduced for the bottleneck nodes.

### 4.9 Retransmission Scenarios for Multi-Beam Nodes

IEEE 802.11 performs a 4-way handshake (RTS, CTS, Data, and ACK) to minimize the collision probability. In Riverbed Modeler, a node extracts the data packet from the buffer queue (at the MAC layer) and stores it in the fragmentation buffer pointer before sending RTS. After receiving CTS, node extracts the data packet from the fragmentation buffer, makes a temporary copy in the wlan_transmit_frame_copy_ptr variable at the function block of the wireless_lan_mac module (which is destroyed upon receiving the ACK), and transmits it. If a beam (at the transmitter node) does not receive a response (i.e., CTS or ACK), it reacquires the channel (using control packets), extracts the data packet from the fragmentation buffer (when the last transmitted frame is RTS) or from the wlan_transmit_frame_copy_ptr variable (when the last transmitted frame is data), and retransmits it.

During CPT, the bottleneck node extracts multiple data packets from their respective locations (i.e., fragmentation buffer(s) or copy pointer(s)) for retransmission. Additionally, if at least one beam achieves successful transmission, the bottleneck node does not consider it a retransmission scenario. The retransmission, therefore, can be categorized into partially unsuccessful and pure retransmission scenarios. In a partially unsuccessful scenario, at least one beam achieves successful transmission, whereas in a pure retransmission scenario, none of the beams receive their desired response. Fig. 8 shows the flowchart to implement both these retransmission scenarios, which is explained below.

We use two global integer variables, *total_desired_frame* and *no_of_frame*, inside the wireless_lan_mac module, to store the number of non-SCH packets transmitted in the last transmission mode and received in the current reception mode, respectively. If *no_of_frame != total_desired_frame* when the frame timeout interrupt is canceled, the bottleneck node knows that it has not received responses on one or more beams, and identifies the beams that need to retransmit their packets. Each unsuccessful beam must update the retry count of its packet and enqueue it at the head of its respective buffer queue so that the packet can be extracted again in the next transmission cycle. For each beam, the node checks whether the last transmission updated its NAV. Recall that the unblocked beam, which transmits the SCH packet, also updates its NAV. Hence, those unsuccessful (and/or unblocked) beams which updated their NAV because of the last transmission reset their NAV to the current time to prevent further delay. The node, then, checks if this beam transmitted a non-SCH packet in the last cycle. If the non-SCH packet was transmitted, the packet's retry count is incremented, and the packet goes back to the head of its queue. At the same time, the node continues its 4-way handshake on its successful beams.

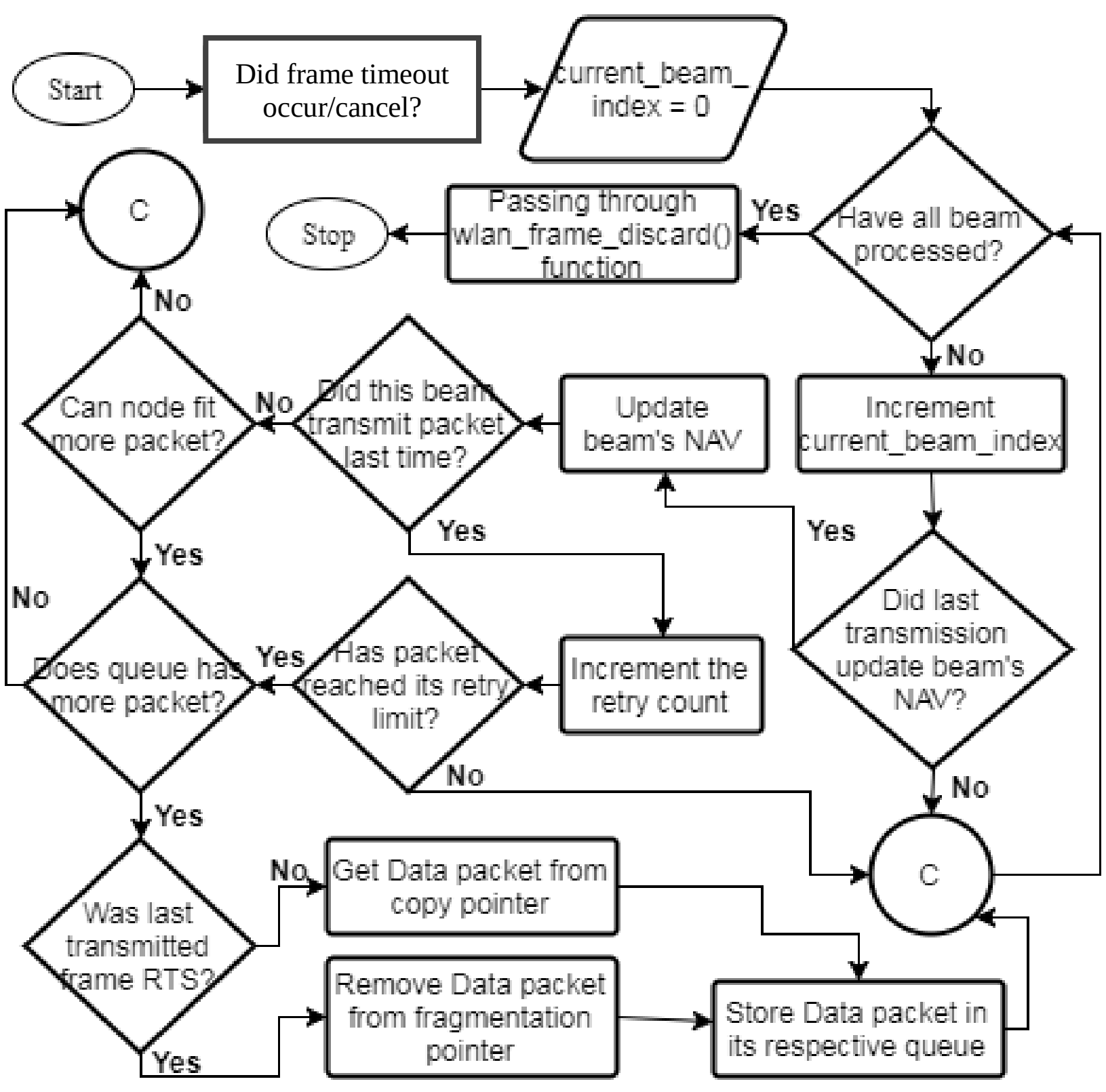

Fig. 8. Flowchart for retransmission scenarios. Here, 'C' is a connector in the flowchart.

In a pure retransmission scenario, the frame timeout interrupt is triggered inside the wait_for_response state which is located inside the process model of wireless_lan_mac module. Here, the bottleneck node compares the *total_desired_frame* variable with its capacity. The maximum number of simultaneous handshakes in which a bottleneck node can be involved is called the *bottleneck capacity*. If the *bottleneck capacity* is more, the bottleneck node can accommodate more packets in its retransmission cycle. This comparison maximizes the utilization of the CPT. After processing all beams, a node calls the wlan_frame_discard() function (located inside the wireless_lan_mac module) to discard the packets which have reached their retry limit. While incrementing the beam's retry count, the node predicts whether this beam's packet is going to be dropped. If yes, the node tries to accommodate a new packet in the retransmission cycle.

If a node decides to accommodate new packets, it sets the global Boolean variable *can_send_more_packets* (introduced inside the process model of the wireless_lan_mac module) to true. At the time of transmission, if this variable is false, the node initiates the retransmission process and refers either to the fragmentation buffer(s) or the wlan_transmit_frame_copy_ptr(s) as discussed earlier. Whereas if the variable is true, the node has at least one new packet which is neither stored in the fragmentation buffer nor in the wlan_transmit_frame_copy_ptr. Hence, the node reinitiates the 4-way handshake on all of its packet-carrying beams.

### 4.10 Support for Multi-Hop Transmission

The MAC layer is responsible for packet forwarding in hop-to-hop communications. However, to analyze MAC performance over a multi-hop scenario, the route information is also needed. This is accomplished by hardcoding the next hop information for each node in the wlan_data_process() function inside the wireless_lan_mac module. When a node receives a data packet, it checks the final destination address (which is address3 field of the IEEE 802.11 data packet header). In Riverbed Modeler, if the node is an access point (AP) and its address is different from the final destination address of the received packet, the MAC stores the packet in the buffer queue for further transmission. However, the packet cannot be stored in the MAC in Riverbed Modeler if the intermediate node is not access point. In order to store the packets at the MAC layer of a non-AP node, we have introduced an OR case to this condition and modified the wlan_hlpk_enqueue() function, which is located inside the wireless_lan_mac module, to take the *queue_index* into consideration (see the pseudocode below).

*if(ap_flag == OPC_BOOLINT_ENABLED* **|| *my_address != final_dest_addr*** *)*

*wlan_hlpk_enqueue (enqueue the packet in the (Remainder(antenna_index,4)+1)$^{th}$ queue)*

Here, the *antenna_index* represents the beam number of the arrived packet. To decide the *queue_index* for the arrived packet, a mapping is needed between beams and queues. In our simulation, each $(i+1)^{\text{th}}$ beam is mapped to $j^{th}$ queue such that:

$$j = (i \; mod \; x) + 1;$$
$$where, x \in \; total \; number \; of \; packet \; queues \; at \; the \; MAC \; layer; \; i \in [0,7] \tag{4}$$

For example, in Fig. 3, when a packet (whose final destination is node 6 in Fig. 9) arrives at *antenna_index* 0 of node 5, it is stored in the first buffer queue at the MAC layer of node 5.

### 4.11 Issue of Multipath

When there is no line of sight (LOS) between the transmitter and receiver, a node may receive multiple copies of the same packet. With multi-beam antenna, a bottleneck node may receive copies of the same packet on its different ports [2,10].

Riverbed Modeler has the wlan_tuple_find() function (defined inside the function block of the wireless_lan_mac module) which checks received packet's transmitter address, destination address, and the sequence number. When a packet traverses on multi-hop paths, a node may receive the same packet on its different beams. To track a packet, Riverbed Modeler attaches a structure, called packet contents, to every packet, which contains fields, such as packet's memory address, format, creation module, creation time, owner, ICI ID, tree ID, etc. Upon receiving the packet, different modules (of the same or a different node) update values of these fields. However, the *tree ID* is assigned only by the source transmitter node. Hence, we use the *tree ID* field of a packet to identify the duplicate packets. The MAC checks the *tree ID* of the packet and compares it with that of the previously processed packet. The packet is processed further only when its *tree ID* is different, otherwise the MAC discards this newly arrived packet.

## 5. SIMULATION SETUP AND RESULTS

We have implemented the multi-beam MAC on the Wireless Suite of Riverbed Modeler in Version 18.0.2. The performance of the multi-beam MAC is evaluated for the concurrent and multi-hop communication scenarios. A 3-hop network topology shown in Fig. 9 is used, where all the nodes use the updated multi-beam node model shown in Fig. 3. The MAC and PHY layer parameters used in our simulations are given in Table 1.

The HPBWs of each antenna in the azimuth and elevation planes are $10^0$, and both the single-beam and multi-beam directional antennas can cover up to 3 km distance. The SRL (short retry limit) for control packets and LRL (long retry limit) for Data packets are 7 and 4, respectively. The DIFS and SIFS durations are 50 µsec and 10 µsec, respectively. The channel bit rate is 1 Mbps per beam (which is the default value in Riverbed Modeler) and the channel fading and noise are ignored. The packet size is 512 Bytes. Buffer size of non-bottleneck (using single beam antenna) and bottleneck (using multi-beam antenna) nodes are 32 kB and 128 kB, respectively. Simulation time for each scenario is 180 s. The *time slot* parameter (see Table 1) represents the total time taken for transmitting one data packet, beginning from channel sensing until the ACK packet is received, when its short and long retry counts are zero (i.e., transmitter node does not retransmit RTS packet).

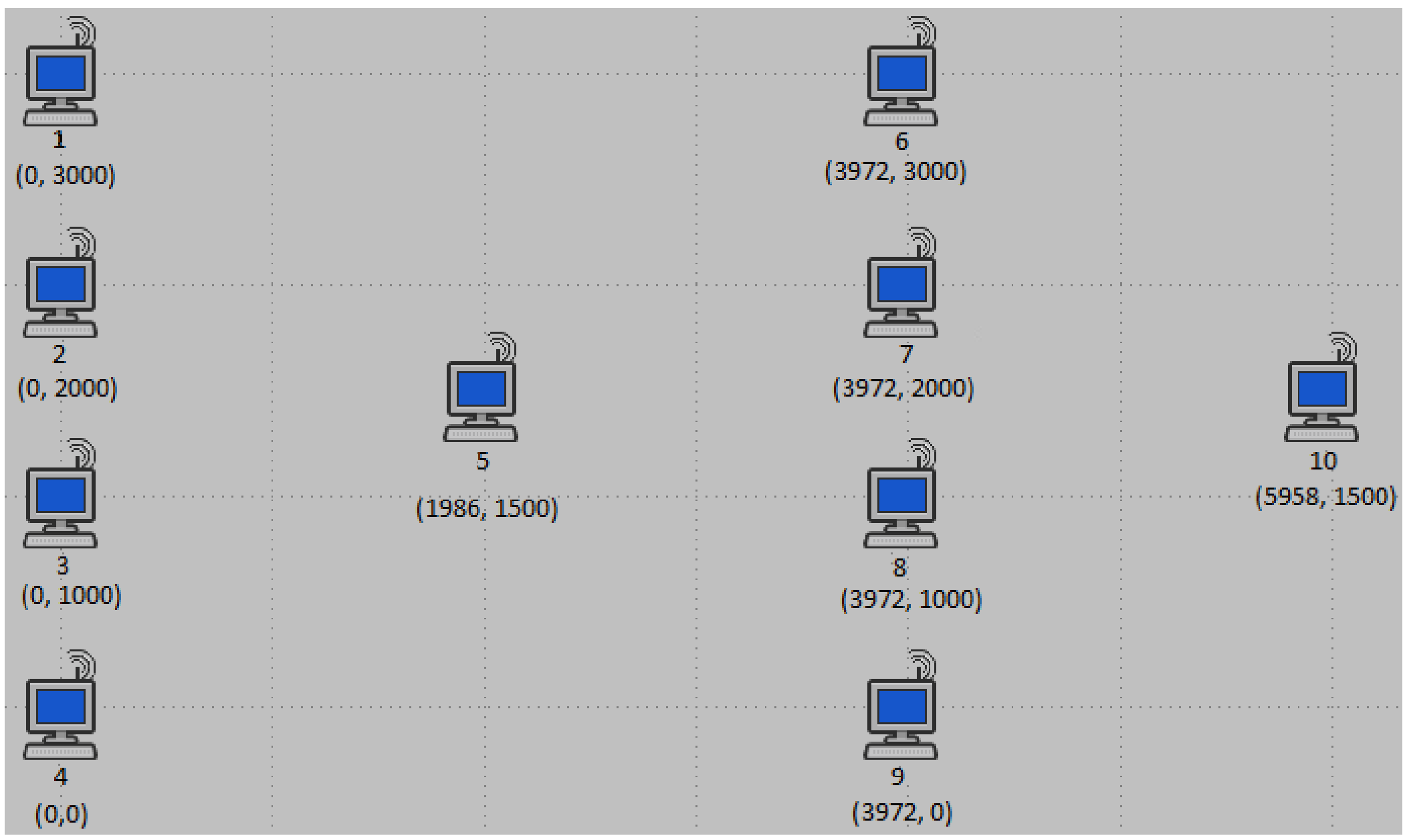

Fig. 9. The 3-hop network topology including the bottleneck nodes 5 and 10.

**Table 1: Simulation Parameters**

| Simulation Parameter | Value | Simulation Parameter | Value | Simulation Parameter | Value |
|---|---|---|---|---|---|
| RTS packet size | 20 Bytes | Buffer size at bottleneck node | 128 kB | Slot size | 20 µs |
| CTS packet size | 14 Bytes | $CW_{min}$ | 15 | DIFS duration | 50 µs |
| ACK packet size | 14 Bytes | $CW_{max}$ | 1023 | SIFS duration | 10 µs |
| Data packet size | 512 Bytes | Short retry limit | 7 | AIFS duration | 20 µs |
| Buffer size at non-bottleneck node | 32 kB | Long retry limit | 4 | Time slot | 4.67 ms |

In Fig. 9, the nodes 5 and 10 are bottleneck nodes, whereas the remaining nodes are non-bottleneck. The bottleneck node 5 has eight beams, pointing towards nodes 1 to 4 and 6 to 9. The bottleneck node 10 uses only four beams which are pointed towards nodes 6 to 9. The beams of the non-bottleneck nodes 1 to 4 are pointed towards the bottleneck node 5. The beam steering delay is assumed to be negligible.

### 5.1 Results for Concurrent Packet Communication

This scenario shows the CPT and CPR capabilities of a bottleneck node. Here, bottleneck node 5 transmits packets to nodes 1 to 4 concurrently on its four different beams (see Fig. 9 for network topology). The Euclidean distance of nodes 2 and 3 to node 5 is 2 km, and nodes 1 and 4 to node 5 is 2.5 km. Similarly, node 10 concurrently receives packets from nodes 6 to 9 on its four beams to test its CPR performance. Here, the Euclidean distance of nodes 7 and 8 to node 10 is 2 km, and nodes 6 and 9 to node 10 is 2.5 km.

Each source generates 250 packets per second (512 Bytes per packet) on each beam with an inter-arrival time of 4 ms at a data rate of 1 Mbps per beam. Since the bottleneck node 5 has one source on each of its four beams, it generates a total of 1000 packets per second. The starting time of its four sources are 10.0 s, 10.000010 s, 10.000020 s, and 10.000030 s, respectively. As described in Section 4.3, a gap is needed in the start times to prevent ICI from getting overwritten. Unlike node 5, nodes 6 through 9 use only one source and hence generate 250 packets per second per node. A non-bottleneck node can store up to 64 packets in its buffer since its buffer size is 32 kB, whereas a bottleneck node can store up to 256 packets in its buffer. The AIFS duration is assumed such that SIFS (10 µs) < AIFS < DIFS (50 µs). Since the slot time is 20 µs, a minimum value of one slot is used for the jump backoff mechanism.

#### *A. Throughput Analysis when Transmitter is a Bottleneck Node*

In this section, we study the throughput performance of the multi-beam MAC scheme for the network topology shown in Fig. 9, where the bottleneck transmitter node 5 is sending data to four receiver nodes 1 through 4, which are located at different distances (nodes 1 and 4 are located at a distance of 2.5 km, whereas nodes 2 and 3 are located at distance of 2.0 km).

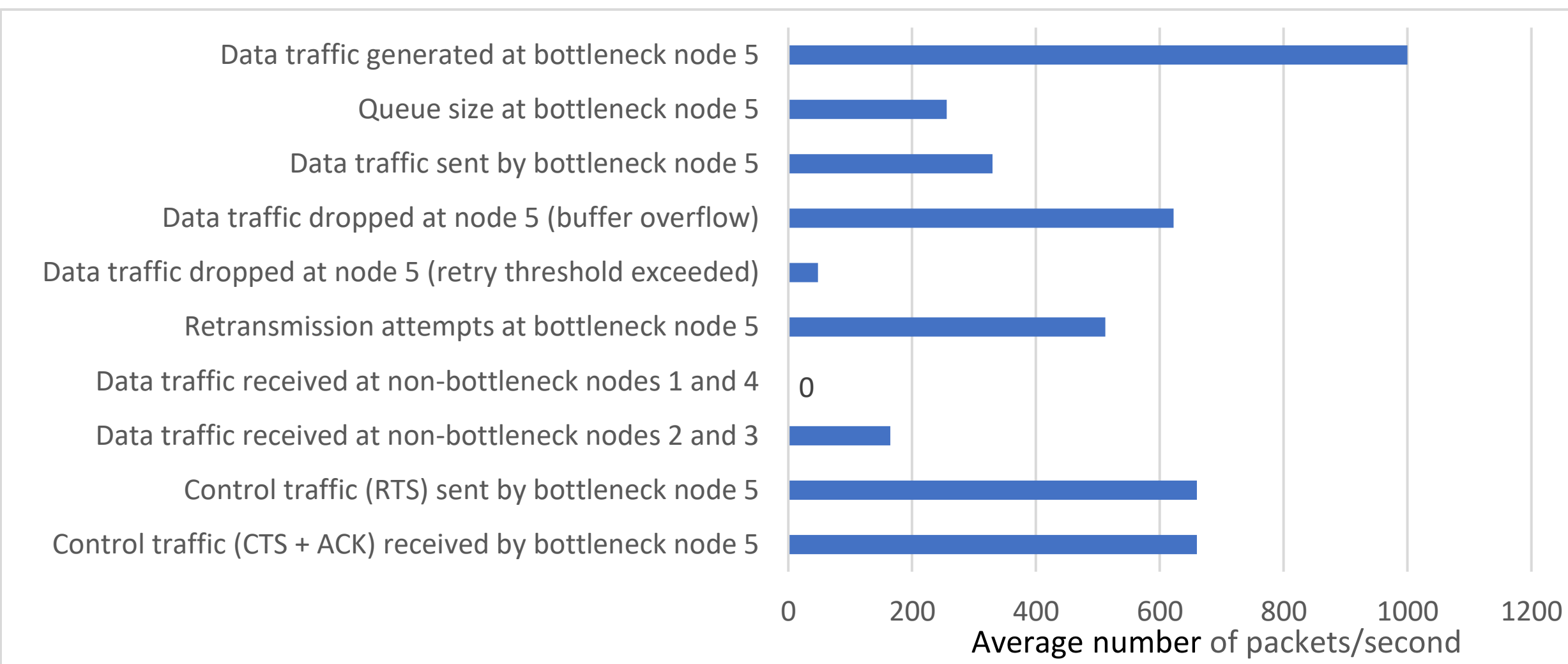

Fig. 10. Data traffic at bottleneck transmitter node 5 and receiver nodes 1-4.

Note that the MAC layer of a bottleneck node processes only those packets which arrive concurrently on its different beams. Although bottleneck node 5 sends RTS packets concurrently on its four beams to nodes 1 to 4, it does not receive four CTS from them at the same time due to its different distances from nodes 1 to 4. As shown in Fig. 10, node 5 transmits 660 RTS packets per second (see 9th bar from top) for nodes 1 through 4 but receives 330 CTS packets per second concurrently only from nodes 2 and 3. As a result, it sends only 330 data packets per second (see 3rd bar from top) out of 1000 packets (see 1st bar from top), with 622 packets per second getting dropped due to the buffer overflow (see 4rth bar from top) and the remaining 48 packets per second getting dropped due to retry threshold exceeded (see 5th bar from top). Node 5 retransmits the packets for nodes 1 and 4 in each communication cycle (see 6th bar from top) and drops them after the maximum retry limit is reached. Since node 5 does not receive any valid RTS or invalid CTS, it does not send any SCH packet. For each data packet transmitted, node 5 receives an ACK. Hence, it receives a total of 660 control packets (CTS+ACK) per second (see the bottom bar). The buffer queue of node 5 is consistently holding 256 packets (see 2nd bar from top). As the retransmission packets occupy the entire buffer space, the MAC layer drops packets coming from the higher layer.

### *B. Throughput Analysis when Transmitter is a Non-Bottleneck Node*

In this section, we study the throughput performance of the multi-beam MAC scheme for the network topology shown in Fig. 9, where four non-bottleneck nodes 6 through 9 transmit packets to the four different beams of the bottleneck receiver node 10. The transmitter nodes are located at different distances from receiver node 10 (nodes 6 and 9 are located at a distance of 2.5 km from node 10, whereas nodes 7 and 8 are located at distance of 2 km).

Since nodes 6 and 9 are farther away from node 10 as compared to nodes 7 and 8 (see Fig. 9), their packets reach node 10 after the packets from nodes 7 and 8. As a result, nodes 6 and 9 increment their node-based backoff and retransmit their packets towards node 10, which reach before the packets from nodes 7 and 8. Therefore, node 10 sets *isValidRTSReceived* variable on its all four beams. It allows bottleneck node 10 to alternatively communicate with node pairs 6 and 9, and 7 and 8, where it transmits CTS packets on two beams (e.g., nodes 7 and 8) and SCH/CTS on the remaining two beams (e.g., node 6 and 9). Fig. 11 shows the data traffic generated at nodes 6 to 9, which is 250 packets per second per node (1st bar). Because of alternate communication, each node transmits only 85 packets per second (see 2nd bar), and the remaining 165 packets per second are dropped (see 3rd bar). Although each node retransmits almost 100 packets per second (see the last bar), no packet is being dropped because of exceeding the retransmission limit. However, the retransmission causes buffer overflow. Similar traffic behavior is also observed for nodes 7, 8 and 9.

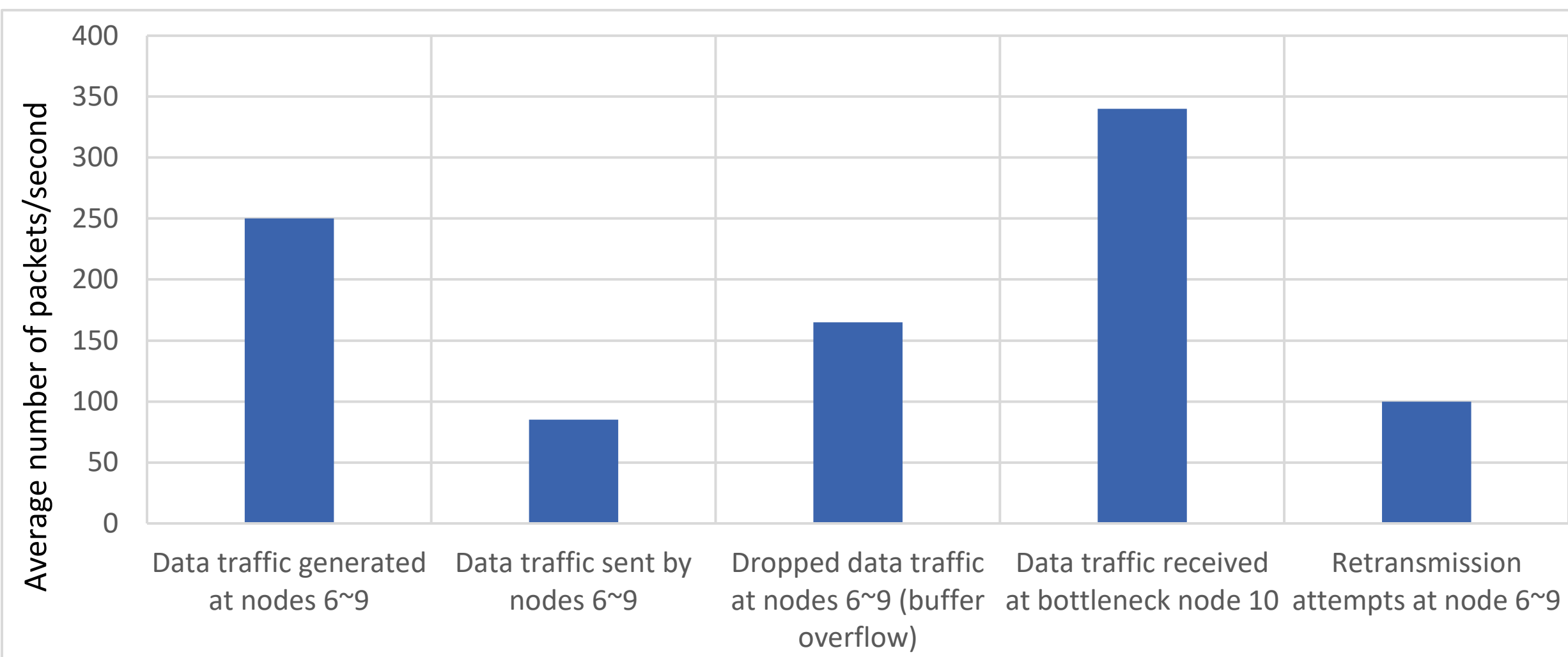

Fig. 11. Data traffic at non-bottleneck transmitter nodes 6~9.

### 5.2 Results for 3-Hop Communication

In Fig. 9, the source nodes 1 through 4 have packets for the destination node 10, which are routed through the bottleneck node 5, followed by nodes 6 through 9. Each source node generates 250 packets per second, where each packet is 512 bytes.

As discussed in Section 5.1B, bottleneck receiver node 5 communicates with node pairs 1 and 4, and 2 and 3 alternatively. Since bottleneck node 5 now has packets to forward, it transmits RTS packets to nodes 6 through 9 and SCH/RTS packets to nodes 1 through 4. Note that nodes 1 through 4 now complete their NAV duration at the same time, and therefore, concurrently transmit their packets to bottleneck node 5. However, bottleneck node 5 only receives packets of nodes 2 and 3 because nodes 1 and 4 are farther away from node 5 as compared to the nodes 2 and 3 (see Fig. 9). Therefore, nodes 1 and 4 fail to send their data packets (shown by 0 in 2nd bar in Fig. 12), whereas 85 packets/sec each from nodes 2 and 3 reach node 5 (see 3rd bar in Fig. 12). As a result, node 10 receives 85 packets/sec each only from nodes 7 and 8 (shown by last bar in Fig. 12).

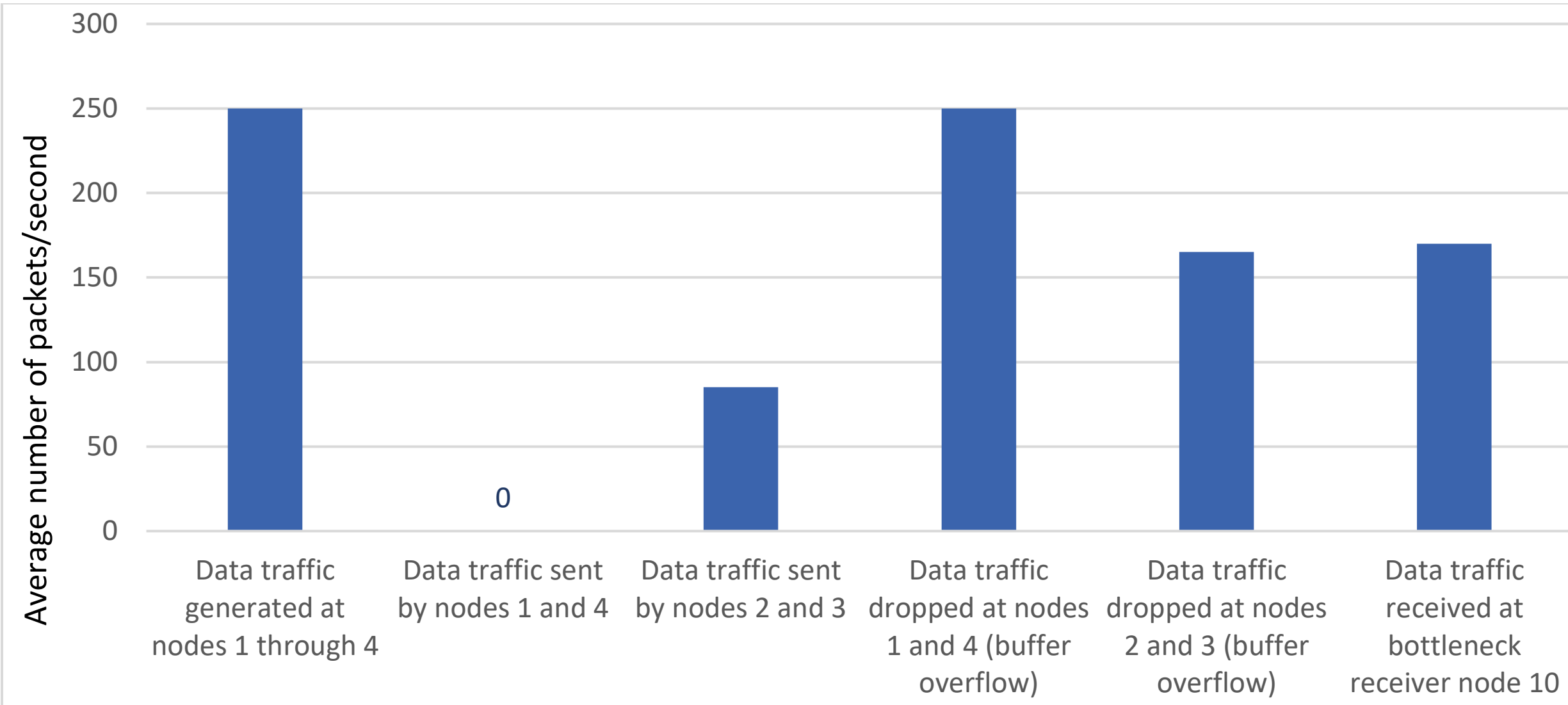

Fig. 12. Data traffic at nodes 1 through 4.

## 6. CONCLUSION

The detailed discussion on the design and implementation of a multi-beam based directional MAC scheme for a multi-hop wireless networks in Riverbed Modeler was presented. Note that the current Riverbed Modeler does not support multi-beam and directional communication. Specifically, the modifications required in the architecture of Riverbed Modeler Wireless Suite for modeling various modules were discussed, such as the multi-beam antenna module (to receive and transmit multiple packets concurrently), multi-beam node module, source and interface modules (to generate and synchronize multiple packets so that the CPT can be performed on a bottleneck node), MAC module (to utilize and optimize the multi-beam capability at the MAC layer), collision avoidance, retransmission, and local node synchronization. These implementations are essential for studying the high-throughput network protocols for single-beam as well as multi-beam directional communication, including the directional multipath routing protocols. This paper would be helpful for researchers who may want to implement and evaluate directional MAC protocols with single-beam as well as multi-beam communication in wireless multi-hop ad hoc networks.

## ACKNOWLEDGEMENT

The authors acknowledge the U.S. government's support in the publication of this paper. This material is based upon the work funded by the U.S. AFRL (Air Force Research Lab), under Contract No. FA8750-14-1-0075. Any opinions, findings and conclusions or recommendations expressed in this material are those of the author(s) and do not necessarily reflect the views of AFRL or the US government. The authors would also like to acknowledge their former group member Nojan Rahimian for his contributions in the initial implementation of directional MAC protocols in Riverbed Modeler.

## REFERENCES

1. O. Bazan, and M. Jaseemuddin, "A survey on MAC protocols for wireless adhoc networks with beamforming antennas," *IEEE Comm. Surveys Tuts.*, vol. 14, no. 2, pp. 216-239, 2012.
2. G. Wang and Y. Qin, "MAC protocols for wireless mesh networks with multi-beam antennas: A survey," *Future of Information and Commun. Conf.*, pp. 117-142, 2019.
3. D. T. C. Wong, Q. Chen, and F. Chin, "Directional medium access control (MAC) protocols in wireless ad hoc and sensor networks: A survey," *J. Sens. Actuator Netw.*, vol. 4, no. 2, pp. 67–153, 2015.
4. J. D. M. M. Biomo, T. Kunz, and M. St-Hilarie, "Exploiting Multi-Beam Antennas for End-to-End Delay Reduction in Ad Hoc Networks," *Mobile Netw. Applications*, pp. 1-13, 2018.
5. S. Say, N. Aomi, T. Ando, and S. Shimamoto, "Circularly multidirectional antenna arrays with spatial reuse based MAC for aerial sensor networks," in *Proc. IEEE Int. Conf. Comm. Workshop*, pp. 2225–2230, 2015.

6. S. Say, H. Inata, and S. Shimamoto, “A hybrid collision coordination-based multiple access scheme for super dense aerial sensor networks,” *IEEE Wireless Comm. and Networking Conf. (WCNC)*, pp. 1-6, 2016.
7. G. Wang, P. Xiao, and W. Li, “A novel MAC protocol for wireless network using multi-beam directional antennas,” *Computing, Networking, and Comm. (ICNC)*, pp. 36-40, 2017.
8. V. Jain, A. Gupta, and Dharma P. Agrawal, “On-demand medium access in multihop wireless networks with multiple beam smart antennas,” *IEEE Trans. Parallel Distrib. Syst.*, vol. 19, no. 4, pp. 489-502, Apr. 2008.
9. I. Stevanovic, A. Skrivervik, and J. R. Mosig, “Smart antenna systems for mobile communications,” [Online]. Available: http://infoscience.epfl.ch/record/140902/files/smart_antennas.pdf.
10. J. Wang, Y. Fang, and D. Wu, “Enhancing the performance of medium access control for WLANs with multi-beam access point,” *IEEE Trans. Wireless Comm.*, vol. 6, no. 2, pp. 556-565, 2007.
11. M. Ascione, G. Bernardi, A. Buonanno, M. D’Urso, M. Felaco, M. G. Labate, G. Prosco, and P. Vinetti, “Simultaneous beams in large phased radar arrays,” *Array Syst. Tech. IEEE Int’l Symp.*, pp. 616-616, 2013.
12. M. Di Filippo, L. Lucci, D. Marabissi, and S. Selleri, “Design of a smart antenna for mobile ad hoc network applications,” *Int’l J. Antennas Propagation*, 2015.
13. R. B. MacLeod, and A. Margetts, “Networked Airborne Communications using Adaptive Multi-Beam Directional Links,” *IEEE Aerosp. Conf.*, pp. 1-7, 2016.
14. H. T. Chou, “An effective design procedure of multibeam phased array antennas for the applications of multisatellite/coverage communications,” *IEEE Trans. Antennas. Propag.*, vol. 64, no. 10, pp. 4218–4227, Oct. 2016.
15. WANSolution Works, “Riverbed Modeler Wireless Suite,” [Online]. Available: https://www.wansolutionworks.com/Modeler.asp.
16. V. Jain, A. Gupta, D. Lal, and D. P. Agarwal, “A cross layer MAC with explicit synchronization through intelligent feedback for multiple beam antennas,” in *Proc. IEEE Globecom ’05*, pp. 3196-3200, 2005.
17. Z.-T. Chou, C.-Q. Huang, and J. M. Chang, “QoS provisioning for wireless LANs with multi-beam access point,” *IEEE Trans. Mobile Computing*, vol. 13, no. 9, pp. 2113-2127, 2014.
18. E. Ulukan and Ö. Gurbuz, “Using switched beam smart antennas in wireless ad hoc networks with angular MAC protocol,” in *Proc. Med-Hoc-Net*, Bodrum, Turkey, Jun. 2004.
19. E. Ulukan and Ö. Gurbuz, “Angular MAC protocol with location based scheduling for wireless ad hoc networks,” in *Proc. IEEE Veh. Tech. Conf. (VTC)*, vol. 3, pp. 1473-1478, May 2005.
20. K. Sundaresan, and R. Sivakumar, “A unified MAC layer framework for ad-hoc networks with smart antennas,” *IEEE/ACM Trans. Netw.*, vol. 15, no. 3, pp. 546-559, Jun. 2007.
21. L. Bao, and J. J. Gracia-Luna-Aceves, “Transmission scheduling in ad hoc networks with directional antennas,” in *Proc. ACM MobiCom ’02*, pp. 4858, 2002.
22. D. Lal, V. Jain, Q.-A. Zeng, and D. P. Agarwal, “Performance evaluation of medium access control for multiple-beam antenna nodes in a wireless LAN,” *IEEE Trans. Parallel Distrib. Syst.*, vol. 15, no. 2, pp. 1117-1129, 2004.
23. L. Kleinrock and F. Tobagi, “Packet switching in radio channels: part I- carrier sense multiple-access modes and their throughput-delay characteristics,” *IEEE Trans. Comm.*, vol. 23, no. 12, pp. 14001416, Dec. 1975.
24. X. Li, L. Hu, K. Bao, F. Hu, and S. Kumar, “Intelligent Multi-Beam Transmission for Mission-Oriented Airborne Networks,” *IEEE Trans. Aerosp. Electron. Syst.*, vol. 55, no. 2, pp. 619-630, 2018.
25. X. Li, F. Hu, J. Qi, and S. Kumar, “Systematic Medium Access Control in hierarchical Airborne Networks with Multi-Beam and Single-beam Antennas,” *IEEE Trans. Aerosp. Electron. Syst.*, vol. 55, no. 2, pp. 706-717, 2018.
26. J. Qi, F. Hu, X. Li, Koushik A M, and S. Kumar, “3-ent (resilient, intelligent, and efficient) medium access control for full-duplex, jamming-aware, directional airborne networks,” *Computer Networks*, vol. 129, Part 1, pp. 251-260, Dec. 2017.
27. J. Qi, X. Li, S. Garg, F. Hu, and S. Kumar, “Riverbed-based network modeling for multi-beam concurrent transmissions,” *Int’l J. Wireless & Mobile Networks*, vol. 9, pp. 1-20, 2017.
28. Ö. Gurbuz, and E. Ulukan, "Wireless LANs with Smart Antennas," [Online]. Available: http://www.sabanciuniv.edu/mdbf/telecom/eng/comnet/cisco/index.htm.
29. Y.-B. Ko, V. Shankarkumar, and N. H. Vaidya, “Medium access control protocols using directional antennas in ad hoc networks,” in *Proc. IEEE INFOCOM*, vol. 1, pp.13-21, March 2000.
30. J. Stine, “Modeling smart antennas in synchronous ad hoc networks using OPNET’s pipeline stages,” *The MITRE Corporation*, Tech. Rep, 2005.
31. C. A. Balanis, Antenna Theory: Analysis and Design. New York, NY, USA: John Wiley & Sons, Inc., 1997.
32. Riverbed SteelCentral, “What does ICI mean and how do you use them when developing OPNET models?,” *Riverbed Support: Knowledge Base*, [Online]. Available: https://supportkb.riverbed.com/support/index?page=content&id=S20734.